\documentclass[usegraphicx,usenatbib,onecolumn]{mn2e}
\usepackage{amssymb}
\usepackage{bm}

\begin{document}
\title{The cross-correlation of the CMB polarisation and the 21~cm line
fluctuations from cosmic reionisation}

\author[Tashiro, H. et al.]
{Hiroyuki Tashiro$^1$, Nabila Aghanim$^1$, Mathieu Langer$^1$, 
\newauthor Marian Douspis$^1$, and Saleem Zaroubi$^2$ \\
  $^1$ Institut d'Astrophysique Spatiale (IAS), B\^atiment 121, F-91405,
Orsay (France);\\
Universit\'e Paris-Sud XI and CNRS (UMR 8617)\\
$^2$ Kapteyn Astronomical Institute, Landleven 12, 
Groningen 9747 AD (The Netherlands)
}

\date{\today}

\maketitle

\begin{abstract}
The cosmic microwave background (CMB) polarisation and the 21~cm line
fluctuations are powerful probes of cosmological reionisation. We
study how the cross-correlation between the CMB polarisation
($E$-modes) and the 21~cm line fluctuations can be used to gain
further understanding of the reionisation history, within the
framework of inhomogeneous reionisation. Since the $E$-mode
polarisation reflects the amplitude of the quadrupole component of the
CMB temperature fluctuations, the angular power spectrum of the
cross-correlation exhibits oscillations at all multipoles.  The first
peak of the power spectrum appears at the scale corresponding to the
quadrupole at the redshift that is probed by the 21~cm line
fluctuations. The peak reaches its maximum value in redshift when the
average ionisation fraction of the universe is about half. On the
other hand, on small scales, there is a damping that depends on the
duration of reionisation. Thus, the cross-correlation between the CMB
polarisation and the 21~cm line fluctuations has the potential to
constrain accurately the epoch and the duration of reionisation.
\end{abstract}


\maketitle

\section{Introduction}

The history of the cosmological reionisation is one of the open
problems of modern cosmology.  Questions like what causes the
reionisation and how it proceeds are intimately related to the
evolution of matter density fluctuations and to the formation of the
first structures \citep{barkana-loeb-2001}. The lack of observational
data makes it quite difficult to answer these questions today. Up to
now, the available probes of the reionisation epoch are just a few,
e.g. the Gunn-Peterson test \citep{gunn-peterson-1965} and the cosmic
microwave background (CMB) radiation, in particular its polarisation
\citep{zaldarriaga-1997}.  The observation of the Gunn-Peterson
effect, which probes the amount of neutral hydrogen, suggests that
full reionisation was accomplished by $z \sim 6$ \citep{Fan-2006}. The CMB
polarisation, in turn, measures the optical depth of the Thomson
scattering from the last scattering surface to the present epoch.  The
WMAP 3 years data provide an optical depth $\tau\sim 0.09$
\citep{page-wmap} suggesting that reionisation began around $z=10$.

In addition to those two probes, the observation of fluctuations of
the 21~cm line background is expected to be one of the most promising
methods to study reionisation \citep{madau-meiksin-rees,
tozzi-madau-2000, ciardi-madau}. The 21~cm line corresponds to the
energy of the hyperfine spin flip transition of neutral hydrogen
atoms.  After decoupling of the baryon temperature from the CMB
temperature around $z\sim 300$, neutral hydrogen atoms absorb or emit
a 21~cm line, depending on their temperature. Since the 21~cm
line is redshifted by the cosmological expansion, we can obtain
redshift slices of the Universe by choosing the frequency of observation.
It is thus possible to follow the evolution of the intergalactic medium before and
during reionisation directly. The 21~cm line observations have
recently gained a lot of attention
\citep{loeb-zaldarriaga-2004,furlanetto-oh-2006,tashiro-21-2006,
cooray-2006} and several projects are designed for measuring the
fluctuating line background (e.g.
MWA\footnote{http://web.haystack.mit.edu/array/MWA},
LOFAR\footnote{http://www.lofar.org},
SKA\footnote{http://www.skatelescope.org}).

In the near future, we will therefore have different direct and
complementary probes of reionisation.  We thus expect that 
cross-correlations between them will provide even more information
than their respective auto-correlations. This is the case especially for
the cross-correlation between the CMB temperature anisotropies and the 21~cm
line fluctuations on large scales 
\citep[$\ell \sim 100$ --][]{alvarez-komatsu,adshead-furlanetto-2008} and
on small scales
\citep[$\ell >1000$
  --][]{salvaterra-ciardi,cooray-2004,slosar-cooray-2007}. Since the
observations of the 21~cm lines are expected to be dominated by
foregrounds and noise \citep{jelic-2008},
calculating the cross-correlation is indeed a good way to extract
information on reionisation.  In particular on large scales, the 21~cm
line fluctuations are correlated with the CMB Doppler temperature
anisotropies which are due to the motion of ionised baryons.
\citet{alvarez-komatsu} showed that the maximum amplitude of the
cross-correlation is reached at the redshift when the ionised fraction
is one half, which potentially provides a way to constrain double
reionisation models \citep[e.g.][]{cen-2003}.

In the present study, we examine the cross-correlation between the CMB
$E$-mode polarisation (a more effective probe than the temperature
anisotropies) and the fluctuations of the 21~cm line background on
large scales.  
Measuring $E$-mode polarisation is one of the present challenging observations
for CMB experiments.
However, projects like \textsc{Planck} and CMB-pol will provide 
precise CMB polarisation maps in the near future.
Similarly, cosmological observation of 21 cm line background is
still belongs to the future, but  
LOFAR and MWA are being completed 
and will soon provide us with maps of neutral hydrogen at high redshifts.
Hence, it is useful to investigate
what information on reionisation we can extract from the
cross-correlation of those maps.  
Indeed, we show here that the cross-correlations
are a potentially powerful tool to constrain the history of cosmic
reionisation.  This article is organised as follows.  
In Sec.~II, we
provide the analytic formula of the cross-correlation between the CMB
$E$-mode polarisation and the fluctuations of the 21~cm line background.
In Sec.~III, we show and discuss the results of the cross-correlation.
Sec.~VI is devoted to the conclusions.  Throughout the paper, we use
WMAP 3-year values for the cosmological parameters, i.e. $h=0.73 \ (H_0=h
\times 100 {\rm km/s / Mpc})$, $T_0 = 2.725$K, $h^2 \Omega _{\rm b}
=0.0223$ and $h^2 \Omega_{\rm m} =0.128$ \citep{spergel-wmap} for a
flat cosmology. We also set the speed of light to $c=1$.

\section{Cross-correlation between CMB $E$-mode polarisation and 21~cm lines}

\subsection{The CMB $E$-mode polarisation}

The CMB polarisation is best described by the Stokes parameters, $Q$ and
$U$.  Since the Stokes parameters have spin 2, they can be decomposed
onto the plane waves and spin-2 spherical harmonics $~_{\pm 2} Y_\ell ^m$
\citep{hu-white},
\begin{equation}
(Q \pm i U)(\eta, {\bm x}, \hat {\bm n}) =
\sum _{\ell\; m}  (-i)^\ell \sqrt{ \frac{4 \pi}{2 \ell +1}}
\int {\frac{d^3 k}{(2 \pi)^3}} \left(E^{(m)}_\ell \pm i
B^{(m)}_\ell\right)
\exp( i {\bm k} \cdot {\bm x}) ~_{\pm 2} Y_\ell ^m (\hat {\bm n}).
\end{equation}
Here, $m=0$, $\pm 1$, $\pm 2$ correspond to the scalar, vector and
tensor types.  In this paper, we only consider $m=0$ modes, because
they dominate other polarisation modes at the scales we are interested
in.
 
The CMB polarisation components, $E^{(0)}_\ell$ and
$B^{(0)}_\ell$, are obtained from the Boltzmann equations for CMB.
According to those, $B^{(0)}_\ell$ =0 for all $\ell$s and the
evolution of $E^{(0)}_\ell$ is given by the following integral
solution,
\begin{eqnarray}
{\frac{E^{(0)}_\ell(\eta_0,k)}{2\ell+1}} =   - {\frac{3}{2}}
\sqrt{\frac{(\ell+2)!}{(\ell-2)!}} 
\int_0^{\eta_0} d\eta  \dot\tau e^{-\tau} 
P^{(0)} \frac{j_\ell(k(\eta_0-\eta))}{(k(\eta_0-\eta))^2},
\label{eq:emode-inte}
\end{eqnarray}
where $P^{(0)}$ is the $m=0$ source term due to Thomson scattering, $\dot
\tau$ is the differential optical depth for Thomson scattering in
conformal time, given by $\dot \tau = n_e \sigma_{\rm T}a$ with the
electron number density $n_e$, the cross section of Thomson scattering
$\sigma_{\rm T}$ and the scale factor $a$ normalised to the present epoch,
and $\tau$ is the optical depth between the conformal times $\eta$ and
$\eta_0$.  The source term $P^{(0)}$ is related to the initial
gravitational potential $\Phi_0$ via
the transfer function $D_E(k, \eta)$,
\begin{equation}
P^{(0)}=D_E(k, \eta ) \Phi_0.
\label{eq:def-transemode}
\end{equation}
We provide the detailed calculation of the transfer function in the
appendix.


We also define the $E$-modes in the direction $\hat {\bm n}$ on the sky
by
\begin{equation}
E( \hat {\bm n}) =
\sum _{\ell\; m}  (-i)^\ell \sqrt{ \frac{4 \pi}{2 \ell +1}}
\int {\frac{d^3 k}{(2 \pi)^3}} E^{(0)}_\ell  
Y_\ell ^m (\hat {\bm n}).
\label{eq:emode-map}
\end{equation}

\subsection{21~cm line fluctuations}

The observed brightness temperature of the 21~cm lines in a direction
$\hat {\bm n}$ and at a frequency $\nu$ is given as in
\cite{madau-meiksin-rees} by
\begin{equation}
T_{21} (\hat {\bm n};\nu) = \frac{\tau_{21}}{(1+z_{\rm obs})}
 (T_{\rm s} -T_{\rm CMB})(\eta_{\rm obs}, \hat {\bm n} (\eta_0-\eta_{\rm
obs})). 
\label{eq:21cmline}
\end{equation}
Here, the conformal time $\eta_{\rm obs}$ refers to the redshift $z_{\rm
obs}$ that satisfies $\nu = \nu_{21}/(1+z_{\rm obs})$ with $\nu_{21}$ being
the frequency corresponding to the 21~cm wavelength.  
The optical depth for the 21~cm line absorption is
\begin{equation}
\tau_{21} 
\sim 8.6 \times 10^{-3} x_{\rm H} \frac{T_{\rm cmb}}{T_{s}}
\left ( \frac{\Omega_{\rm b} h^2}{0.02} \right)
\left[
\left ( \frac{0.15}{\Omega_{\rm m} h^2} \right)
\left ( \frac{1+z_{\rm obs}} {10} \right)
\right] ^{1/2},
\label{eq:tau21}
\end{equation}
where $x_{\rm H}$ is the fraction of neutral hydrogen, and $T_{\rm s}$
is the spin temperature given by the ratio of the number
density of hydrogen atoms in the excited state to that of hydrogens in the
ground state.

Equations (\ref{eq:21cmline}) and (\ref{eq:tau21}) show that the observed
brightness temperature of the 21~cm lines will reflect 
 the baryon density fluctuations $\delta _{\rm b}$ and 
the fluctuations of the ionised fraction $\delta_{\rm x}$. We can rewrite
Eq.~(\ref{eq:21cmline}) in the linear approximation for both
 $\delta _{\rm b}$ and $\delta _x$ as
\begin{equation}
T_{21} (\hat {\bm n} ;\nu) = 
[1- \bar x_e(1+ \delta_x)](1+\delta_b) T_{0},
\label{eq:21cmlinefl}
\end{equation}
where we $x_e = 1- x_{\rm H}$ is the ionised fraction and
\begin{equation}
T_0 =23 
\left ( \frac{\Omega_{\rm b} h^2}{0.02} \right)
\left[
\left ( \frac{0.15}{\Omega_{\rm m} h^2} \right)
\left ( \frac{1+z_{\rm obs}}{10} \right)
\right] ^{1/2}
{\rm mK}.
\end{equation}
Here, we assume that $T_s$ is much larger than the CMB temperature
which is valid soon after the beginning of the reionisation
\citep{ciardi-madau}.

After performing the Fourier transform of Eq.~(\ref{eq:21cmlinefl}),
we obtain the 21~cm line fluctuations,
\begin{eqnarray}
\delta T_{21}(\hat {\bm n};\nu ) &\equiv&
T_{21} (\hat {\bm n} ;\nu)-T_0
\nonumber \\
&=& T_{0} \int \frac{dk ^3}{(2 \pi)^3}
\left [(1- \bar x_e) \delta_{\rm b} - \bar x_e \delta_x
\right ]
\exp{[- i {\bm k} \cdot ((\eta -\eta_{\rm obs}) \hat {\bm n})]}.
\label{eq:fluc-21}
\end{eqnarray}
Therefore, the 21~cm line fluctuation map at the frequency $\nu$ 
can be described by
\begin{equation}
\delta T_{21}( \hat {\bm n};\nu ) = T_{0} \sum_\ell
\int \frac{dk^3}{(2 \pi)^3}\sqrt{4 \pi (2 \ell +1)}  
\left [(1- \bar x_e) (1+ F \mu^2)\delta_{\rm b} - \bar x_e \delta_x
\right ]
j _\ell(k (\eta_0-\eta_{\rm obs})) Y_\ell ^0( \hat{\bm n}).
\label{eq:21cm-map}
\end{equation}
where we applied Rayleigh's formula to Eq.~(\ref{eq:fluc-21}),
\begin{equation}
\exp (-i {\bm k} \cdot {\bm x})=
\sum_{\ell} \sqrt{4 \pi (2 \ell +1)} (-i)^\ell j_\ell(kr) Y_\ell ^0(
\hat{\bm n}),
\end{equation}
where $\hat{\bm n}$ is measured in the coordinate system where $\bm
k$ is $\hat {\bm e}_3$.  In Eq.~(\ref{eq:21cm-map}), we included the factor
$(1+ F \mu^2)$ to account for the enhancement of the fluctuation amplitude due to the redshift distortion (Kaiser effect \citep{kaiser-1987}) on the
21~cm line fluctuations where $\mu = \hat {\bm k} \cdot \hat {\bm n}$ and
$F=d \ln g / d \ln a$ with $g(a)$ the linear growth factor of baryon
fluctuations \citep{bharadwaj-ali}.


\subsection{Cross-correlation between 21~cm and $E$-mode fluctuations}\label{subsec:cross}

The angular power spectrum is defined as the average of the spherical
harmonic coefficients $a _{\ell m}$ over the $(2 \ell +1)$ $m$-values,
\begin{equation}
C_\ell = \sum _m \frac{\langle |a _{\ell m} |^2  \rangle}{(2 \ell +1)},
\end{equation}
where the $a _{\ell m}$ are defined for an arbitrary sky map $f( {\bm \hat
  n})$ as  
\begin{equation}
f( {\bm \hat n}) = \sum_{\ell m} a _{\ell m} Y_\ell ^m.
\end{equation}
From Eqs.~(\ref{eq:emode-map}) and (\ref{eq:21cm-map}), we can thus
derive the angular power spectrum of the cross-correlation between
$E$-modes and 21~cm line fluctuations 
\begin{eqnarray}
C_\ell ^{E-21} &=&  4 \pi T_0\int d \eta \int \frac{dk^3}{(2 \pi)^3 }
\frac{\left \langle  
E_\ell ^0 (\eta, k) \left [
 (1- \bar x_e) (1+ F \mu^2) \delta_{\rm b} - \bar x_e \delta_x \right]
\right  \rangle }{(2 \ell +1)} j_\ell (k( \eta_0 -\eta_\nu))
\nonumber \\
&=&
 -\frac{3}{\pi} T_0 \sqrt{\frac{(\ell+2)!}{(\ell-2)!}} 
\int {dk } \int d \eta k^2 \dot \tau e^{-\tau} 
D_E  (k, \eta )
\left[ \frac{4}{3} 
(1- \bar x_e) P_{\Phi \delta_{\rm b}}- \bar x_e P_{\Phi \delta_x}
\right] \frac{j_\ell (k( \eta_0 -\eta_{\rm obs})) j_\ell  (k( \eta_0
-\eta))}{(k( \eta_0 -\eta))^2},
\label{eq:cross-cl}
\end{eqnarray}
where $P_{\Phi \delta_{\rm b}}$ and $P_{\Phi \delta_x}$ are the power
spectra of the cross-correlation between the gravitational potential
and the baryon density fluctuations, and between the gravitational
potential and the fluctuations of the ionised fraction, respectively.
In the second line, we use $F \langle \mu^2 \rangle = 1/3$ for the
matter dominated epoch.

The cosmological linear perturbation theory provides the relation
between the baryon density fluctuations and the initial gravitational
potential \citep[e.g.][]{kodama-sasaki-1984}, $\delta_{\rm b}(k, \eta) =
k^2
D_{\rm b} (k, \eta) \Phi_0(k)$ where $D_{\rm b}$ is the transfer
function of baryons. 
The power spectrum $P_{\Phi \delta_{\rm b}}$ can thus be written in terms
of the initial power spectrum of the gravitational potential $P_{\Phi}$ as
\begin{equation}
P_{\Phi \delta_{\rm b}} = k^2 D_{\rm b}(k, \eta ) P_{\Phi}.
\end{equation}
Although the reionisation process is not well-known, we can reasonably
expect that there is a relation between the fluctuations of the ionised
fraction and that of the matter density. Ionising sources are formed
in dense regions. They ionise the surrounding medium with an efficiency
that depends on the density of the medium. This translates into two
possible cases: One where ionised fluctuations and matter over-densities
coincide, and one where ionised fluctuations and matter density are
anti-biaised \citep[e.g.][]{benson-2001}. Following
\citet{alvarez-komatsu}, we assume that the fluctuations of the ionised
rate are associated with the matter density contrast using the
Press-Schechter description \citep[][]{ps-1974}.  As a result,
the power spectrum $P_{\Phi
\delta_x}$ is given by
\begin{equation}
\bar x_e P_{ \Phi \delta_x} =
- \bar x_{\rm H} \ln \bar x_{\rm H} 
[\bar b_{\rm h}-1-f] D_{\rm m}(k, \eta ) k^2 P_{\Phi},
\end{equation}
where $D_{\rm m}$ is the transfer function of matter (both dark and
baryonic), $\bar b_{\rm h}$ is the average bias of dark matter halos
more massive than $M_{\rm min}$, the minimum mass of the source of
ionising photons
\begin{equation}
\bar b_{\rm h} = 1 + \sqrt{\frac{2}{\pi}} \frac{e^{-\delta_c ^2 /2
\sigma^2(M_{\rm min})}}{f_{\rm coll} \sigma (M_{\rm min})},
\end{equation}
where $\sigma(M)$ is the variance of the density fluctuations smoothed
with a top-hat filter of the scale corresponding to a mass $M$, and
$f_{\rm coll}$ is the fraction of matter collapsed into halos with
$M>M_{\rm min}$.  In this paper, we choose $M_{\rm min}$ such that the
halo virial temperature is $T_{\rm vir}(M_{\rm min}) = 10^4$ K. This
choice corresponds to the assumption that the ionising sources form in dark
matter halos where the gas cools efficiently via atomic cooling. 
The parameter $f$ describes the reionisation regime we
are interested in. For $f=0$, we are in the ``photon-counting limit''
case where recombinations are not important and where the progress of
the reionisation depends on the number of ionising photons only.  The
over-dense regions contain more collapsed objects which are sources of
ionising photons.  Therefore, in this case, ionisation in 
over-dense regions is easier than in  under-dense regions.  On the
contrary, $f=1$ indicates the ``Str\"{o}mgren limit" case where 
ionisation is balanced by recombination.  Although the
over-dense regions contain more sources of ionising photons, the
recombination rate in  over-dense regions is higher than in
under-dense regions. Hence,  over-dense regions in the $f=1$ case
have a lower ionised fraction than in the $f=0$ case
\citep[for details, see][]{alvarez-komatsu}.

Finally, in order to calculate Eq.~(\ref{eq:cross-cl}), we need the
evolution of the mean neutral hydrogen fraction for which we use a
simple parameterisation based on two key quantities, the reionisation
epoch (defined as the redshift at which the ionised fraction equals
0.5), $z_{\rm re}$, and the duration of reionisation, $\Delta z$,

\begin{equation}
\bar x _{\rm H} (z) = \frac{1}{1 + \exp[-(z-z_{\rm re})/\Delta z]}.
\label{eq:xh}
\end{equation} 

In the following we will vary both $z_{\rm re}$ and $\Delta z$, and
investigate their effect on the cross-correlation signal.


\section{Results and discussions}

As one can see in Eq. (\ref{eq:cross-cl}), the angular power spectrum
$C_\ell^{E-21}$ of the cross-correlation is a function of several key
parameters that encode the prominent details of the reionisation
history. Consequently, and as we show here, the global shape of the
$C_\ell^{E-21}$ as well as the amplitude and redshift dependence of
its first peak are features that can be traced back to the
reionisation characteristics.

First it is worth noting that because of the double integration of the
spherical Bessel functions in Eq.~(\ref{eq:cross-cl}), the angular
power spectrum reflects the source term $P(k_{\rm obs})$, namely the
quadrupole term of the CMB temperature anisotropy, at $z_{\rm obs}$
where $k_{\rm obs}$ satisfies $k_{\rm obs} = \ell / (\eta_0 -
\eta_{\rm obs})$.  Accordingly, the angular power spectrum of the
cross-correlation exhibits its first peak at a multipole $\ell<10$
which corresponds to the angular separation of the quadrupole at
$z_{\rm obs}$. The oscillations at higher $\ell$s ($>10$) are due to the
free streaming of the quadrupole at redshifts higher than $z_{\rm
obs}$.

\begin{figure}
  \begin{tabular}{cc}
   \begin{minipage}{0.5\textwidth}
  \begin{center}
    \includegraphics[keepaspectratio=true,height=50mm]{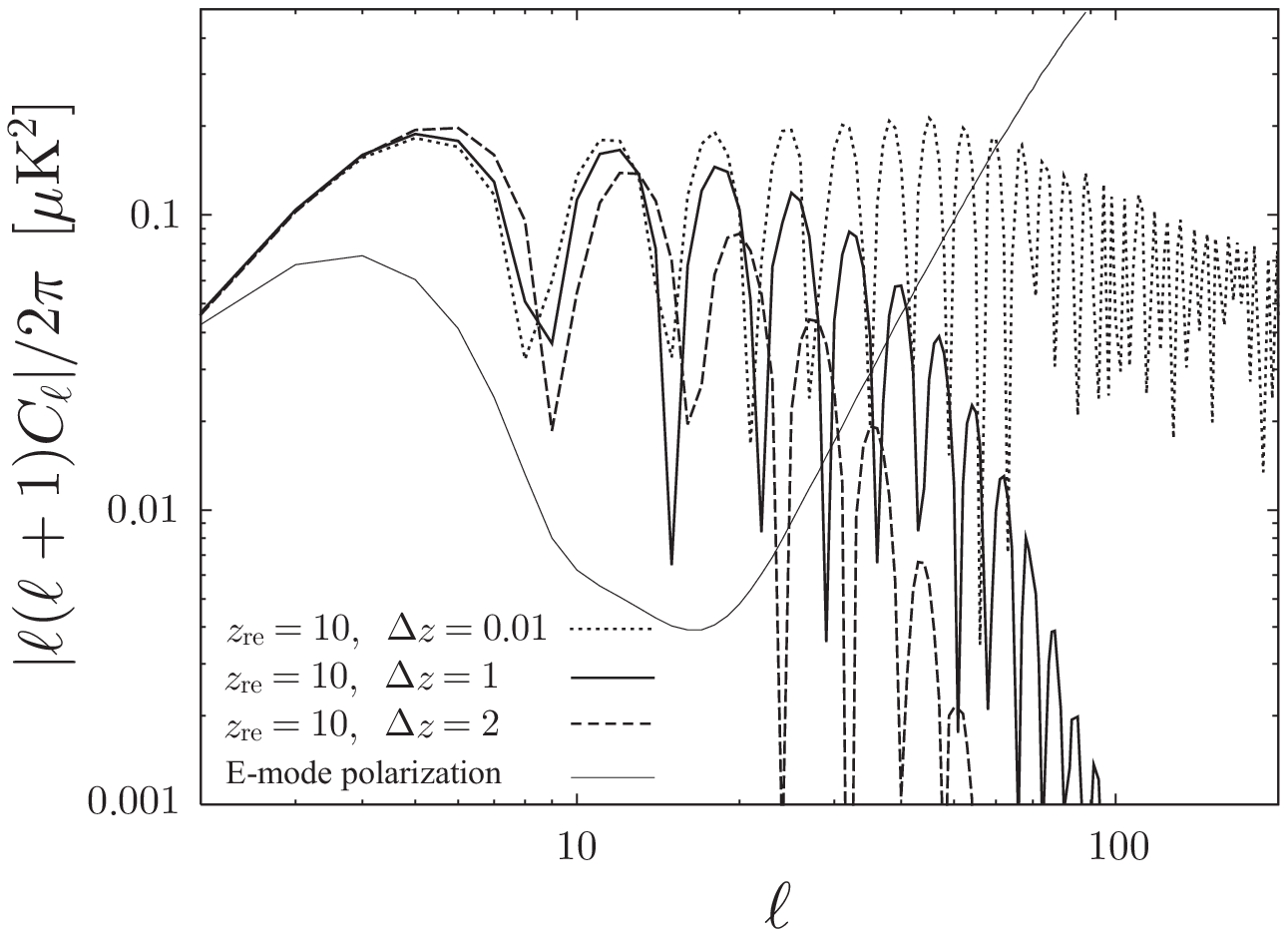}
  \end{center}
  \end{minipage}
   \begin{minipage}{0.5\textwidth}
  \begin{center}
    \includegraphics[keepaspectratio=true,height=50mm]{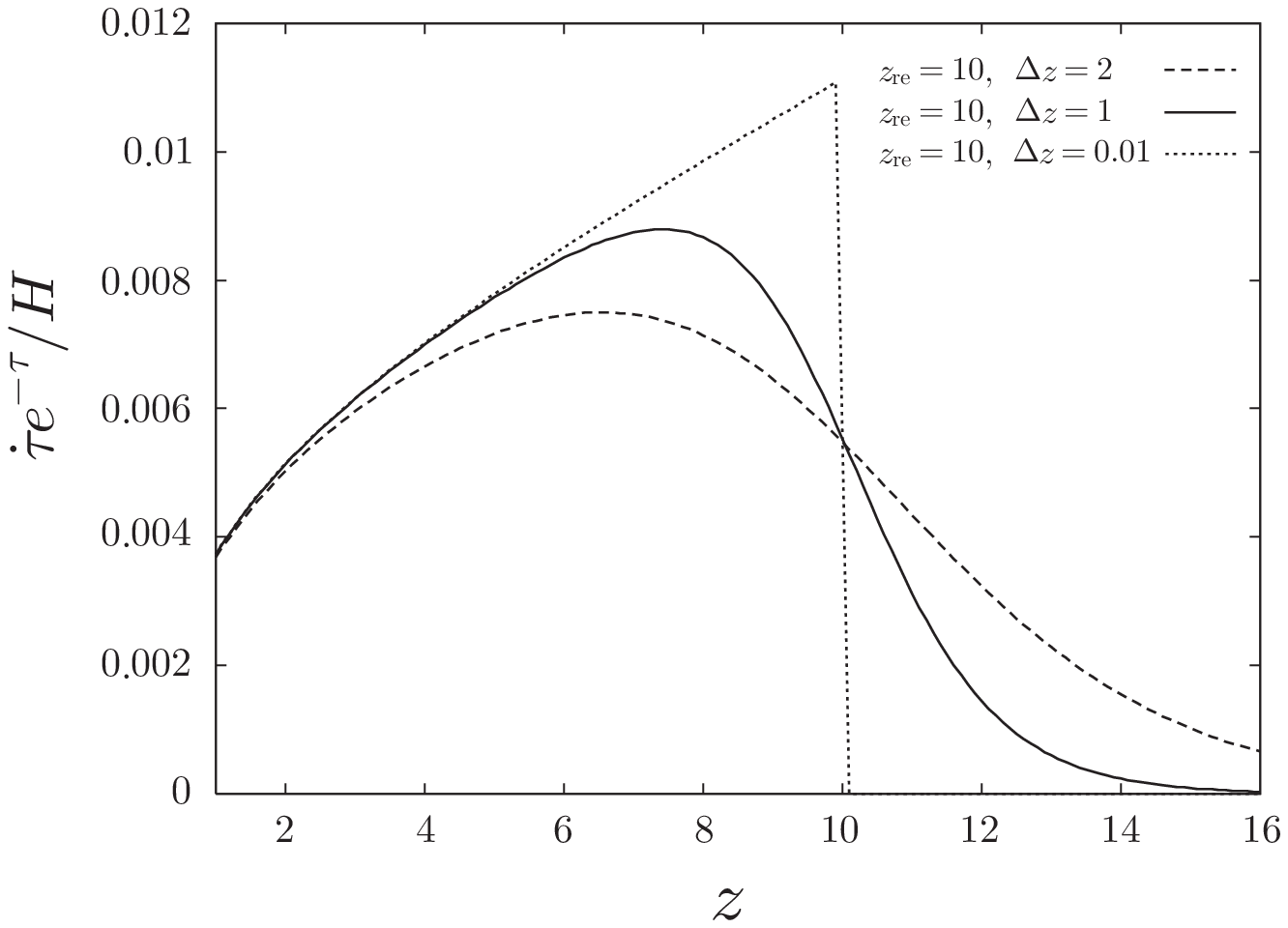}
  \end{center}
   \end{minipage}
  \end{tabular}
  \caption{The left panel shows the angular power spectra of the
  cross-correlation between $E$-mode polarisation and 21~cm line
  fluctuations.  We set $z_{\rm re} = 10$, $z_{\rm obs} = 10$ and
  $f=0$ for all curves.  The dotted line represents the angular
  spectrum for $\Delta z=0.01$.  The solid and dashed lines are for
  $\Delta z=1$ and $\Delta z=2$, respectively. For comparison the thin
  solid line, shows the angular power spectrum of the CMB $E$-modes for
  $z_{\rm re} = 10$ and $\Delta z=1$.  The right panel is the
  evolution of the visibility function $\dot \tau e^{-\tau}$ in the
  same reionisation model as the left panel. The dotted, solid and
  dashed lines are for $\Delta z=0.01$, $\Delta z=1$ and $\Delta z=2$,
  respectively.}
  \label{fig:clre10}
\end{figure}

We first investigate the effect of the duration of reionisation
$\Delta z$. We present in the left panel of Fig.~\ref{fig:clre10}
three cases: One corresponding to a instantaneous reionisation $\Delta
z=0.01$ and two with longer durations ($\Delta z=1$ and 2). We compute
the angular spectra of the cross-correlation between 21~cm line
fluctuations and the CMB $E$-mode fluctuations for those different
durations but for the same given epoch of reionisation, $z_{\rm re}
=10$, an observing redshift, $z_{\rm obs}=10$, and in the case of the
``photon-counting limit'' ($f=0$).  The difference in the duration of
reionisation induces both a minor and a major effects on the spectra.
The minor effect is the observed shift of the peak position to higher
$\ell$s with increasing durations.  Since a longer duration increases
the contribution of the visibility function $\dot \tau e^{-\tau}$ in
Eq.~(\ref{eq:cross-cl}), which we plot in the right panel of
Fig.~\ref{fig:clre10}, the contributions from redshifts larger than
$z_{\rm obs}$ accumulate in the calculation of the
integration. However, spherical Bessel functions have a steep cutoff
after $\eta_{\rm obs}$ so that the contributions from epochs earlier
than $\eta_{\rm obs}$ are negligible.  Therefore, additional effective
contributions come from only the former epoch $\eta < \eta_{\rm obs}$,
and make the peak shift toward higher $\ell$ modes.

The major effect of $\Delta z$ exhibited in the plot (left panel of
Fig.~\ref{fig:clre10}) is the damping at high $\ell$s with increasing
durations of reionisation.  This occurs again due to the increase of
the amplitude of the visibility function at $z > z_{\rm re}$. At high
$\ell$s, the source term $P^{(0)}$ exhibits rapid oscillations. Since
the tail of the visibility function is larger, the range of the
integral is larger, and it thus encompasses more oscillations. As a
result, many more cancellations occur which damp the overall power. We
note that the damping depends on the duration of the reionisation
which is an advantage of observing the cross-correlation signal.  As a
matter of fact, it is difficult to see differences among CMB
polarisation spectra for different durations, because the Thomson
optical depth is not strongly sensitive to this parameter.  However
the plots in the left panel of Fig. \ref{fig:clre10}, clearly show
that the damping of angular spectrum of the cross-correlation is
sensitive to the duration of reionisation making it a potential tool
for constraining this parameter of the reionisation history.


\begin{figure}
  \begin{center}
    \includegraphics[keepaspectratio=true,height=50mm]{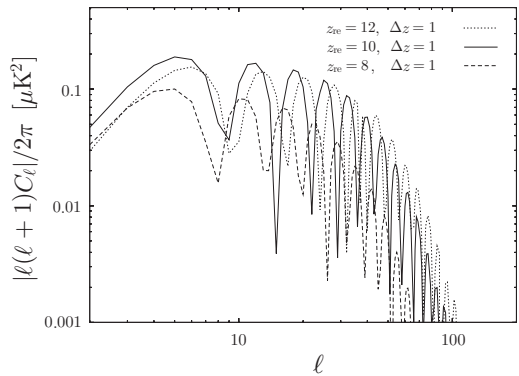}
  \end{center}
  \caption{The angular power spectra of the cross-correlation between
  $E$-mode polarisation and 21~cm line fluctuations.  The dotted line
  is for $z_{\rm}=12$. The solid and dashed lines are for $z_{\rm}=10$
  and $z_{\rm}=8$, respectively.  We use $\Delta z =1$, $z_{\rm
  obs}=10$ and $f=0$ for all curves.}
  \label{fig:differentrecross}
\end{figure}

We now vary the reionisation redshift $z_{\rm re}$ keeping the other
parameters fixed, and setting $z_{\rm obs} = 10$, $\Delta z =1$ and
$f=0$.  We find that increasing $z_{\rm re}$ shifts the position of
the first peak to small scales (Fig.~\ref{fig:differentrecross}). The
reason is the same as the one invoked for the shifts observed in the
left panel of Fig.~\ref{fig:clre10}.  When reionisation happens early,
the integration of the $E$-mode polarisation in
Eq.~(\ref{eq:cross-cl}) takes into account more contribution from
higher redshifts. This induces a shift of the peak position to higher
$\ell$ modes.

\par\bigskip

\begin{figure}
  \begin{center}
    \includegraphics[keepaspectratio=true,height=50mm]{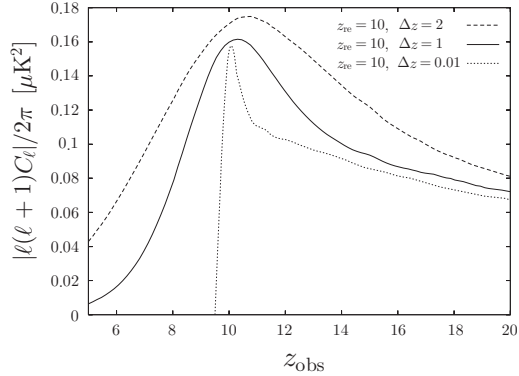}
  \end{center}
  \caption{The amplitude of the first peak of the cross-correlation
  signal as a function of redshift $z_{\rm obs}$. The dotted, solid
  and dashed lines indicate the amplitudes for $\Delta z=0.01$,
  $\Delta z=1$ and $\Delta z=2$, respectively. For all plots, we set $z_{\rm
  re} = 10$ and $f=0$.}
   \label{fig:peakzzre10}
\end{figure}

Figure \ref{fig:peakzzre10} shows the evolution of the first peak
amplitude as a function of $z_{\rm obs}$, the redshift of the 21~cm
line emission, for the three above-mentioned durations of reionisation
$\Delta z$.  Here, we set $z_{\rm re} =10$ and $f=0$.  As shown by
\citet{alvarez-komatsu} using Eq. (\ref{eq:xh}), the amplitude of the
cross-correlation of the CMB Doppler temperature anisotropies and the
21~cm line fluctuations reaches its maximum value at $z_{\rm re}=10$
where the ionised fraction is about half. The same is true in the case
of the cross-correlation of 21~cm line fluctuations and $E$-mode
polarisation.  The reason for this is that the CMB polarisation is
produced in the ionised regions whereas 21~cm line fluctuations are
associated with neutral regions.  We note that the signal in
Fig. \ref{fig:peakzzre10} is smaller at lower redshifts than at higher
redshifts.  This is due to the fact that the 21~cm signal vanishes
once reionisation is completed, i.e. at redshifts smaller than $z_{\rm
re}$. For higher redshifts ($z_{\rm obs}>z_{\rm re}$), the low
scattered quadrupole component of the $E$-modes is suppressed like in
the lower redshifts part, but the integration in the $E$-modes picks
up contributions from the reionisation epoch and the $E$-modes are
correlated with a larger contribution of the 21~cm signal. As a
result, the amplitude of the cross-correlation does not decline as
steeply as for $z_{\rm obs}<z_{\rm re}$.
The difference between the curves is easily interpreted. In the
$\Delta z=2$ case, full reionisation takes longer to complete than in
the $\Delta z=1$ case. Therefore, neutral regions persist longer,
broadening the skewed bell-shape of the curve centered on $z_{\rm obs}
= z_{\rm re}$.  The slight difference in the peak value is due to the
cumulative effect of the integral in Eq. (\ref{eq:cross-cl}). This is
particularly well illustrated by the case of ``instantaneous
reionisation'' ($\Delta z=0.01$).  It is worth noting that our
assumption, $T_{s} \gg T_{\rm cmb}$, is formally not valid before the
beginning of reionisation. In order to evaluate the cross-correlation
signal correctly, we should rather take into account the full
evolution of $T_{s}$ with redshift in the calculation of the 21~cm
line fluctuations. However, by the time the ionisation fraction has
risen above the per cent level, efficient Compton and X-ray heating
will guarantee that the spin temperature be well above $T_{\rm cmb}$
\citep[e.g.][]{prit-fur}.



\begin{figure}
  \begin{center}
    \includegraphics[keepaspectratio=true,height=50mm]{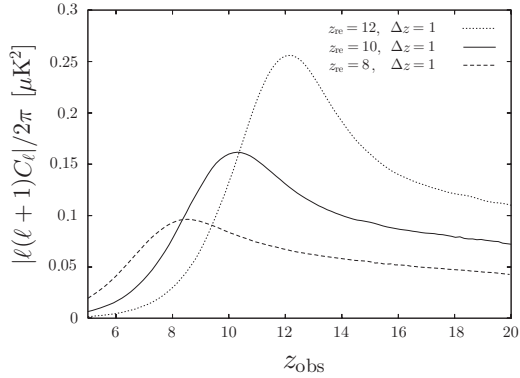}
  \end{center}
  \caption{The evolution of the first peak amplitude for different
  $z_{\rm re}$.  The dotted, solid and dashed lines represent the
  evolution for $z_{\rm re}=12$, $z_{\rm re}=10$ and $z_{\rm re}=8$,
  respectively.  In this figure, we assume $\Delta z =1$ and $f=1$.}
    \label{fig:peakzzred}
\end{figure}

The evolution of the first peak amplitude for different $z_{\rm re}$ is
shown in Fig.~\ref{fig:peakzzred}. As expected, the amplitude reaches its
maximum value at each $z_{\rm re}$.  Since the 21~cm optical depth is
higher at high $z_{\rm obs}$, the amplitude of the 21~cm line
fluctuations at higher redshifts is higher than that at lower
redshifts.  Therefore, and as exhibited by the comparison of the
amplitudes for the different $z_{\rm re}$, the maximum value of the
first peak cross-correlation amplitude is larger in the early
reionisation case than in the late reionisation case.  This again is
the same evolution that shows up in the power spectrum of the
cross-correlation between the CMB temperature anisotropy and the 21~cm
line fluctuations \citep[cf. Fig.  2 of][]{alvarez-komatsu}.  The
evolution, with the observed redshift, of the shape and the amplitude
of the first peak of cross-correlation between the 21~cm line
fluctuations and $E$-mode fluctuations should allow us to constrain
both the duration and redshift of reionisation.

\begin{figure}
  \begin{tabular}{cc}
   \begin{minipage}{0.5\textwidth}
  \begin{center}
    \includegraphics[keepaspectratio=true,height=50mm]{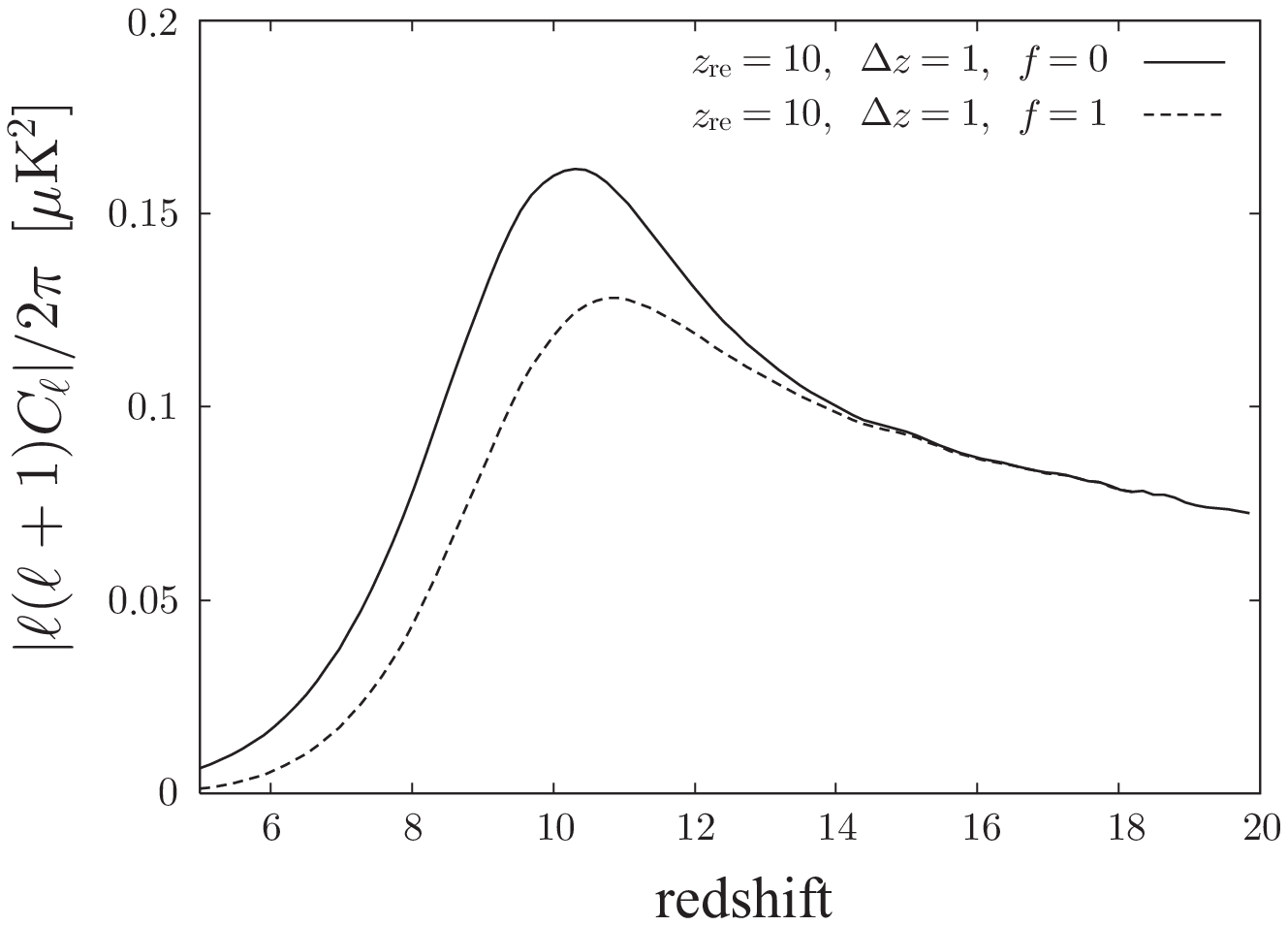}
  \end{center}
  \end{minipage}
   \begin{minipage}{0.5\textwidth}
  \begin{center}
    \includegraphics[keepaspectratio=true,height=50mm]{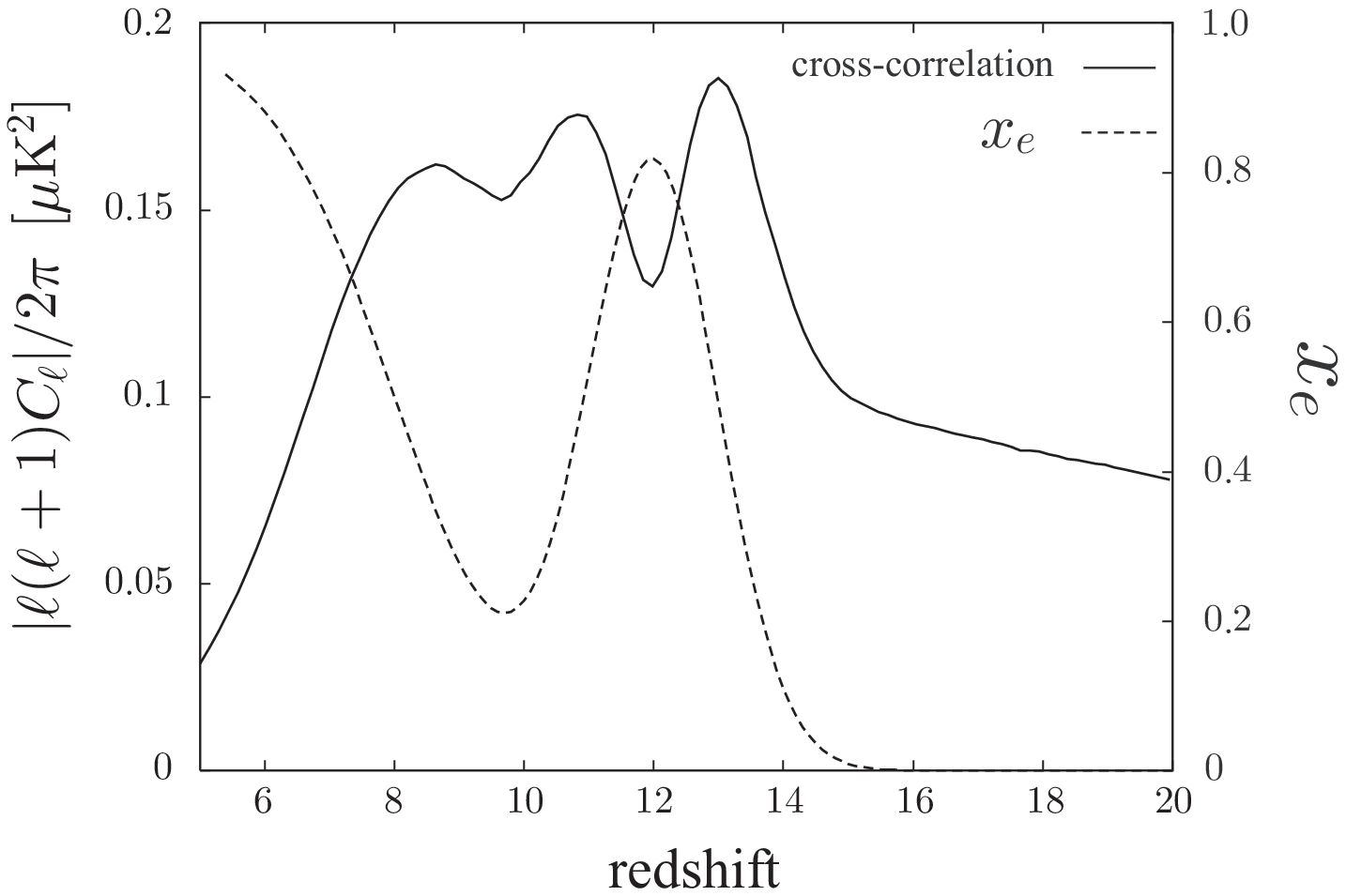}
  \end{center}
   \end{minipage}
  \end{tabular}
  \caption{The left panel shows the evolution of the first peak
amplitude for two extreme values of the parameter $f$.  The solid and
dashed lines are for $f=0$ and $f=1$, respectively. We set $z_{\rm
re}=10$ and $\Delta z = 1$ in all plots.  The right panel shows the
evolution of the first peak amplitude in the case of a double
reionisation model.  The solid line (left axis) is for the first peak
amplitude at given redshift.  The dotted line (right axis) indicates
the ionisation fraction in the double reionisation model.}
  \label{fig:peakzzreionize}
\end{figure}

Ultimately, the parameter $f$ is not independent of the epoch $z_{\rm
re}$ of reionisation and its duration $\Delta z$. However, making that
dependency explicit would require the very detailed description of the
reionisation processes that we are still lacking today. We can
nevertheless illustrate how the cross-correlation power spectrum
changes for different global reionisation scenarios.  The left panel
in Fig.~\ref{fig:peakzzreionize} shows the variation of the evolution
of the first peak amplitude with the parameter $f$.  There is no
difference of the peak position in both spectra for the ``photon
counting limit''$f=0$ and the ``Str\"{o}mgren limit'' $f=1$ but the
amplitudes are different.  When $f$ increases, recombinations become
more and more important. Accordingly, over-dense regions in the case
of $f=0$ have higher ionisation fraction than in the case of $f=1$ as
mentioned in Sec.~\ref{subsec:cross}.  Since the fluctuations of the
neutral fraction are defined by $\delta_{\rm H} = -\delta_x$, the
absolute value of the 21~cm line fluctuations for $f=0$ is higher than
for $f=1$.  Therefore, the absolute value of the amplitude of the CMB
$E$-mode anisotropies and the 21~cm line fluctuations correlation for
$f=0$ is higher than for $f=1$ as shown in the left panel of
Fig.~\ref{fig:peakzzreionize}.

Finally, we also calculate the cross-correlation between the 21~cm line
fluctuations and the $E$-mode CMB anisotropies in the case of a double
reionisation.  
 For illustrative purposes, we chose a simple parametrisation of the
mean ionised fraction
\begin{equation}
 \bar{x}_e(z) = 1- \frac{1}{ 1+\exp[-(z-z_{\rm re})/ \Delta z]}
+  A  \exp \left[-\frac{(z-z_{\rm first})^2 }{  \Delta z_{\rm
first}^2}\right],
\end{equation}
which is the simple reionisation case formula with
$z_{\rm re} =8$ and
$\Delta z =1$ on top of which we added a Gaussian with $A=0.8$, $z_{\rm
first} =12$, and $\Delta z_{\rm first}^2=2$ that encodes a first passage
through an eighty per cent reionised universe at a redshift of twelve.
These parameters have been adjusted so as to reproduce the same value of the
optical depth ($\tau=0.08$) as the single reionisation model shown in the
left panel of Fig. ~\ref{fig:peakzzreionize}. The redshift evolution of
$\bar{x}$ in this toy model is represented
by the dashed line in the right panel of Fig.~\ref{fig:peakzzreionize}
(read right axis for values of $\bar{x}_e$). In the same figure, the solid
line shows the corresponding evolution of the first peak of the
cross-correlation (left axis).  Every time the
ionised fraction takes the value 0.5 there is a peak in the
cross-correlation signal. In the case of our model for double
reionisation, we observe three peaks, i.e. the ionisation fraction
crosses the 0.5 line thrice. However, the two peaks at $z_{\rm obs}
\sim 8$ and $11$ are very close to each other.  If two consecutive
crossings of the 0.5 line occur very soon after each other, the
corresponding peaks in the cross-correlation signal will be
essentially indistinguishable. 
Similarly to what has been
found by \citet{alvarez-komatsu}, this behaviour is a very interesting
property as it implies that we can use the ``trajectory'' of the
evolution of the peak amplitude to constrain the reionisation model.

\section{Conclusion}

We have examined the cross-correlation between the $E$-mode CMB
polarisation and the fluctuations of the 21~cm line background.  In
particular, we calculated the angular power spectrum of the
cross-correlation on large scales using a simple parametrisation of
reionisation, focusing explicitly on the redshift $z_{\rm re}$ of its
occurrence and on its duration $\Delta z$. We also considered the case
of double reionisation for which we obtained specific signatures.

The angular power spectrum of the cross-correlation traces the
quadrupole component of the CMB temperature anisotropies at the redshift
where the 21~cm line is observed.  Both the amplitude and the position
of the first peak depend on the reionisation epoch, as well as on the
redshift $z_{\rm obs}$ probed by the 21~cm observations. We showed
that the signal peaks when $z_{\rm obs}$ matches $z_{\rm re}$, in the
very same way as does the cross-correlation between the CMB Doppler
temperature anisotropies and the fluctuations of the 21~cm line background
\citep[see][]{alvarez-komatsu}, that is when the 21~cm observations
probe the epoch at which the ionised fraction becomes one half. When
probing higher redshifts, the polarisation and 21~cm cross-correlation
signal decreases slowly, whereas it falls off rapidly when lower
redshifts $z_{\rm obs}$ are considered.

The location of the first peak in the ($z_{\rm obs}$ -- $C^{E-21}_\ell$)
plane is essentially independent of $\Delta z$.  Therefore, tracking
the evolution of the first peak amplitude by scanning the sky in
$z_{\rm obs}$ bins would allow to determine $z_{\rm re}$. Moreover,
the width of the bell-shaped curve describing the evolution with
$z_{\rm obs}$ of the $C_\ell^{E-21}$ first peak depends on $\Delta
z$. The duration of the reionisation is thus also accessible through
this observable quantity.

We also showed that, in addition to the amplitude and the position of
the first peak, the duration of reionisation induces a significant
damping of the oscillations on small scales. This strong dependence on
the duration shows specifically in the cross-correlation power
spectrum as it does not arise in the auto-correlations power spectra.

These significant features, which are the oscillations and the maximum 
at the redshift when the ionised fraction are one half,
is explained by Eq.~(\ref{eq:cross-cl}) which is an exact analytic expression.
They are robust and independent of the details of the model.
Assumptions are only introduced to relate the gravitational potential
to the baryon density contrast and to the ionised fraction in Eq.~(\ref{eq:cross-cl}).
For the $\phi-\delta_b$ relation, we assume linear perturbation theory 
which is valid on large scales ($\ell > 100$) and at high redshift ($z \sim 10$).
For the $\phi$-$\delta_x$ relation, we need to assume a reionisation model. 
We choose a simple model based on the halo model.
In our redshift range when the reionised fraction is around one half,
as shown in many numerical studies (e.g., \citet{iliev-mellema-2006}),
the typical size of ionised bubbles is smaller than 
the instrumental resolution.
In this case, we can assume that 
the ionised fraction can be set by the number of ionising sources
(for a detailed discussion see the appendix in \citealt{alvarez-komatsu}).
Numerical simulations that 
provide detailed descriptions of the reionisation history 
can test those assumptions.

In the present study, we assumed the observation of the 21~cm line
ideal. In practice, it amounts to consider that the frequency window
function is the delta function $\delta (\nu -\nu_{\rm obs})$, implying
that observations are done at the specified frequency $\nu_{\rm obs}$
only.  Of course, for real observations, the finite instrumental
frequency width has to be taken into account. For too broad window
functions, this is susceptible to induce an additional damping of the
higher $\ell$ oscillations. We thus checked this effect in the case of
the LOFAR interferometer that has a frequency width of 4 MHz. As a
result, we did not obtain any additional significant damping of the
oscillations for $\ell < 200$.

Both CMB and 21 cm signals suffer from the foreground contaminations.
Some of them will be correlated like the synchrotron emission from our Galaxy.
Recent studies have shown that 
the cleaning of foreground were under control in both 
CMB and 21 cm observations \citep{bock-2006,jelic-2008}.
In addition, the 
multi frequency information on the CMB
will allow the cross-correlation signal to be
even less sensitive to the correlated foregrounds.

Finally, due to the cosmic variance that limits CMB measurements on
large scales, it may actually be difficult to pin-point the first peak
with great accuracy. Moreover, ground-based telescopes designed for
observations of the 21~cm line background cannot cover the whole
sky. Those two facts are not important limitations for our study,
since the damping of oscillations in the cross-correlation power
spectrum on intermediate scales ($10<\ell<100$), which are well enough
sampled, depends strongly on the reionisation evolution.
Actually, even if the reionisation duration is so long that 
we cannot detect the oscillations due to the strong damping,
it is still possible to obtain a lower limit for the duration.
We will study this in more detail with numerical simulations in the future.

%


\appendix
\section{The CMB $E$-mode polarisation}

The source term of $E$-mode polarisation in Eq.~(\ref{eq:emode-inte}) \citep{hu-white},
\begin{equation}
P^{(0)}(\eta,k) = \Theta_2^{(0)}(\eta,k)  -\sqrt{6} E_2^{(0)}(\eta,k),
\end{equation}
where $\Theta_2^{(0)}$ is the quadrupole moment of the CMB temperature
anisotropies.

The CMB temperature anisotropies are evaluated by the following integral solutions
obtained from the Boltzmann equations,
\begin{equation}
\frac{\Theta_\ell^{(0)}(\eta_0,k) }{ 2\ell + 1}
=
\int_0^{\eta_0} d\eta  e^{-\tau} \sum_{\ell'} 
  S_{\ell'}^{(0)}(\eta)  j_\ell^{(\ell' )}(k(\eta_0-\eta)),
\label{eq:temp-inte}
\end{equation} 
where the $j_\ell^{(\ell )}$ are expressed in terms of the spherical Bessel
functions $j_\ell(x)$, 
\begin{equation}
j_\ell^{(0)}(x) = j_\ell(x)   ,\qquad 
j_\ell^{(1)}(x) = j_\ell'(x)  ,\qquad 
j_\ell^{(2)}(x) = \frac{1}{2} [3 j_\ell''(x) + j_\ell(x) ].  
\label{eq:def-jl}
\end{equation}
and $S_{\ell}^{(0)}$ are source terms written as
\begin{equation}
S_0^{(0)} = \dot\tau \Theta_0^{(0)} - \dot \Phi  , 	\qquad 
S_1^{(0)} = \dot\tau v_B + k\Psi  , 		\qquad 
S_2^{(0)} = \frac{\dot\tau}{10} \left[ \Theta_2^{(0)}  -
\sqrt{6} E_2^{(0)} \right],
\label{eq:source-temp}
\end{equation}
where $v_b$ is the velocity of the baryon fluid,
$\Phi$ and $\Psi$ are the scalar-type metric perturbations
in the conformal Newtonian gauge. 
The cosmological linear perturbation theory provides
the evolution of $\Phi$ and $\Psi$ and the initial condition of $\Theta_0$ 
in the matter dominated epoch as \citep{kodama-sasaki-1984}
\begin{equation}
\Psi = - \Phi = - \frac{9}{10} \Psi (0),
\qquad
\Theta_0 (0)= \frac{3}{5} \Psi (0),
\label{eq:metric-ini}
\end{equation}
where $\Psi (0)$ is the initial metric perturbation.

Before recombination, photons and baryons are tightly coupled by 
Compton scattering. Hence, the tight-coupling approximation is a strong
tool for solving Eqs.~(\ref{eq:emode-inte}) and (\ref{eq:temp-inte}).
\citet{hu-sugiyama-1994} provided a useful analytic expression of the
monopole component in this approximation,
\begin{equation}
\Theta_0 (\eta) + \Phi (\eta)
=[\Theta_0 (0) + \Phi (0)] \cos k r_{\rm s} (\eta) 
-\frac{k}{\sqrt{3}} \int ^\eta _0 d \eta' 
[\Phi (\eta')-\Psi (\eta')] \sin[k r_{\rm s}(\eta)-k r_{\rm s}(\eta ')],
\label{eq:acoustic-tight}
\end{equation}
where $r_{\rm s}$ is the integration of the sound speed $c_s = 1 /\sqrt{3(1+R)}$,
\begin{equation}
r_{\rm s}= \int d \eta \frac{1}{\sqrt{3 (1+R)}}.
\end{equation}
The dipole and quadrupole components, and the baryon velocity are written as
\begin{equation}
\Theta_1 = -\frac{3}{k} ( \dot \Theta_0 +\dot \Phi),
\qquad
\Theta_2 = \frac{ 4 \sqrt{4}}{9} \frac{ k }{\dot \tau} \Theta_1,
\qquad
v_{\rm B} = \Theta_1.
\label{eq:multi-tight}
\end{equation}
Accordingly, the quadrupole component of $E$-modes is given by
\begin{equation}
E_2 = - \frac{\sqrt{6}}{4} \Theta_2.
\label{eq:emode-tight}
\end{equation}

After recombination, the tight-coupling approximation is no longer valid.
Using Eqs.~(\ref{eq:metric-ini})-(\ref{eq:emode-tight}) as initial
conditions,
we solve Eqs.~(\ref{eq:emode-inte}) and (\ref{eq:temp-inte}) numerically 
in order to obtain the transfer function $D_E(k, \eta )$ in Eq.~(\ref{eq:def-transemode}).


\begin{thebibliography}{99}

\bibitem[\protect\citeauthoryear{{Adshead} \& {Furlanetto}}{{Adshead} \&
  {Furlanetto}}{2008}]{adshead-furlanetto-2008}
{Adshead} P.,  {Furlanetto} S.,  2008, MNRAS, 384, 291

\bibitem[\protect\citeauthoryear{{Alvarez}, {Komatsu}, {Dor{\'e}} \&
  {Shapiro}}{{Alvarez} et~al.}{2006}]{alvarez-komatsu}
{Alvarez} M.~A., {Komatsu} E., {Dor{\'e}} O., {Shapiro} P.~R., 2006,
ApJ, 647, 840

\bibitem[\protect\citeauthoryear{{Barkana} \& {Loeb}}{{Barkana} \&
 {Loeb}}{2001}]{barkana-loeb-2001}
{Barkana} R.,  {Loeb} A.,  2001, Phys. Rep., 349, 125

\bibitem[\protect\citeauthoryear{{Benson}, {Nusser}, {Sugiyama} \&
{Lacey}}{{{Benson} et~al.}}{2001}]{benson-2001}
{Benson} A. J., {Nusser} A., {Sugiyama} N., {Lacey} C. G., 2001,
MNRAS, 320, 153

\bibitem[\protect\citeauthoryear{{Bharadwaj} \& {Ali}}{{Bharadwaj} \&
  {Ali}}{2004}]{bharadwaj-ali}
{Bharadwaj} S., {Ali} S.~S., 2004, MNRAS, 352, 142


\bibitem[\protect\citeauthoryear{{Bock}, {Church}, {Devlin}, {Hinshaw}, G. 
\& et al.}{{Bock} et~al.}{2006}]{bock-2006}
{Bock} J.,  et al.,  2006, astro-ph/0604101


\bibitem[\protect\citeauthoryear{{Cen}}{{Cen}}{2003}]{cen-2003}
{Cen} R., 2003, ApJ, 591, 12



\bibitem[\protect\citeauthoryear{{Ciardi} \& {Madau}}{{Ciardi} \&
  {Madau}}{2003}]{ciardi-madau}
{Ciardi} B.,  {Madau} P.,  2003, ApJ, 596, 1

\bibitem[\protect\citeauthoryear{{Cooray}}{{Cooray}}{2004}]{cooray-2004}
{Cooray} A.,  2004, Phys. Rev. D, 70, 063509

\bibitem[\protect\citeauthoryear{{Cooray}}{{Cooray}}{2006}]{cooray-2006}
{Cooray} A.,  2006, Physical Review Letters, 97, 261301

\bibitem[\protect\citeauthoryear{{Fan}, {Strauss}, {Becker}, {White}, {Gunn} \&
  {Knapp}}{{Fan} et~al.}{2006}]{Fan-2006}
{Fan} X.,  et al.,  2006, AJ, 132, 117

\bibitem[\protect\citeauthoryear{{Field}}{{Field}}{1958}]{field}
{Field} G., 1958, Proc. IREE Aust., 46, 240


\bibitem[\protect\citeauthoryear{{Furlanetto}, {Oh} \&
  {Pierpaoli}}{{Furlanetto} et~al.}{2006}]{furlanetto-oh-2006}
{Furlanetto} S.~R.,  {Oh} S.~P.,    {Pierpaoli} E.,  2006, Phys. Rev. D, 74,
  103502

\bibitem[\protect\citeauthoryear{{Gunn} \& {Peterson}}{{Gunn} \&
  {Peterson}}{1965}]{gunn-peterson-1965}
{Gunn} J.~E.,  {Peterson} B.~A.,  1965, ApJ, 142, 1633


\bibitem[\protect\citeauthoryear{{Hu} \& {Sugiyama}}{{Hu} \&
  {Sugiyama}}{1995}]{hu-sugiyama-1994}
{Hu} W.,  {Sugiyama} N.,  1995, ApJ, 444, 489

\bibitem[\protect\citeauthoryear{{Hu} \& {White}}{{Hu} \&
  {White}}{1997}]{hu-white}
{Hu} W.,  {White} M.,  1997, Phys. Rev. D, 56, 596

\bibitem[\protect\citeauthoryear{{Iliev}, {Mellema}, {Pen} \& et al.}{{Iliev} et~al.}
{2006}]{iliev-mellema-2006}
{Iliev}, I.~T., {Mellema}, G., {Pen}, U.-L., {Merz}, 
H., {Shapiro}, P.~R., {Alvarez}, M.~A., 2006
MNRAS, 369, 1625


\bibitem[\protect\citeauthoryear{{Jelic} et al.}{{Jelic}
  et~al.}{2008}]{jelic-2008}
{Jelic} V.,  et~al.,  2008, ArXiv 0804.1130


\bibitem[\protect\citeauthoryear{{Kaiser}}{{Kaiser}}{1987}]{kaiser-1987}
{Kaiser} N., 1987, MNRAS, 227, 1

\bibitem[\protect\citeauthoryear{{Kodama} \& {Sasaki}}{{Kodama} \&
  {Sasaki}}{1984}]{kodama-sasaki-1984}
{Kodama} H.,  {Sasaki} M.,  1984, Progress of Theoretical Physics Supplement,
  78, 1

\bibitem[\protect\citeauthoryear{{Loeb} \& {Zaldarriaga}}{{Loeb} \&
  {Zaldarriaga}}{2004}]{loeb-zaldarriaga-2004}
{Loeb} A.,  {Zaldarriaga} M.,  2004, Physical Review Letters, 92, 211301

\bibitem[\protect\citeauthoryear{{Madau}, {Meiksin} \& {Rees}}{{Madau}
  et~al.}{1997}]{madau-meiksin-rees}
{Madau} P.,  {Meiksin} A.,    {Rees} M.~J.,  1997, ApJ, 475, 429

\bibitem[\protect\citeauthoryear{{Page}, {Hinshaw}, {Komatsu}, {Nolta},
  {Spergel}, {Bennett} \& {Barnes}}{{Page} et~al.}{2007}]{page-wmap}
{Page} L., et~al.,  2007, ApJS, 170, 335

\bibitem[\protect\citeauthoryear{{Press} \& {Schechter}}{{Press} \&
{Schechter}}{1974}]{ps-1974}
{Press} W.H., {Schechter} P., 1974, ApJ, 187, 425

\bibitem[\protect\citeauthoryear{{Pritchard} \& {Furlanetto}}{{Pritchard}
\& {Furlanetto}}{2007}]{prit-fur}
{Pritchard} J. R., {Furlanetto} S. R., 2007, MNRAS, 376, 1680

\bibitem[\protect\citeauthoryear{{Salvaterra}, {Ciardi}, {Ferrara} \&
  {Baccigalupi}}{{Salvaterra} et~al.}{2005}]{salvaterra-ciardi}
{Salvaterra} R.,  {Ciardi} B.,  {Ferrara} A.,    {Baccigalupi} C.,  2005,
  MNRAS, 360, 1063

\bibitem[\protect\citeauthoryear{{Slosar}, {Cooray} \& {Silk}}{{Slosar}
  et~al.}{2007}]{slosar-cooray-2007}
{Slosar} A.,  {Cooray} A.,    {Silk} J.~I.,  2007, MNRAS, 377, 168

\bibitem[\protect\citeauthoryear{{Spergel}, {Bean}, {Dor{\'e}}, {Nolta},
  {Bennett} \& {Dunkley}}{{Spergel} et~al.}{2007}]{spergel-wmap}
{Spergel} D.~N., et~al.,  2007, ApJS, 170, 377

\bibitem[\protect\citeauthoryear{{Tashiro} \& {Sugiyama}}{{Tashiro} \&
  {Sugiyama}}{2006}]{tashiro-21-2006}
{Tashiro} H.,  {Sugiyama} N.,  2006, MNRAS, 372, 1060

\bibitem[\protect\citeauthoryear{{Tozzi}, {Madau}, {Meiksin} \& {Rees}}{{Tozzi}
  et~al.}{2000}]{tozzi-madau-2000}
{Tozzi} P.,  {Madau} P.,  {Meiksin} A.,    {Rees} M.~J.,  2000, ApJ, 528, 597

\bibitem[\protect\citeauthoryear{{Wouthuysen}}{Wouthuysen}{1952}]{wout}
{Wouthusen} S., 1952, AJ, 57, 31

\bibitem[\protect\citeauthoryear{{Zaldarriaga}}{{Zaldarriaga}}{1997}]
{zaldarriaga-1997}
{Zaldarriaga} M.,  1997, Phys. Rev. D, 55, 1822

\end{thebibliography}

%

\end{document}